\documentclass[authoryear,review,3p,times]{elsarticle}
\usepackage{amssymb}
\usepackage{amsmath}
\usepackage{lineno}
\usepackage{bm} 
\usepackage{longtable}
\usepackage{caption}
\usepackage{soul}
\usepackage{color}
\usepackage{gensymb}
\usepackage{hyperref}
\usepackage{nomencl} % If needed
\makenomenclature

\graphicspath{{figures/}}

\usepackage{array}

\journal{Annals of Nuclear Energy}

\usepackage{multicol}
\usepackage{nomencl} % If needed
\makenomenclature

\begin{document}

\begin{frontmatter}


\title{Adaptive Control for a Physics-Informed Model of a Thermal Energy Distribution System: Qualitative Analysis}

%%% Authors (use arabic numbers: 1, 2, 3, etc. for affiliationNumber)
%%% \addAuthor{GivenName MiddleInitial. FamilyName}{affiliationNumber}
\author[1]{Paul Seurin}
\ead{paul.seurin@inl.gov}
\author[2]{Anuradha Annaswamy}
\author[1]{Linyu Lin}

%%% Affiliations (from authblk)
%%% \addAffiliation{affiliationNumber}{Name of Institute, City, State/Country}
\affiliation[1]{Idaho National Laboratory, 1955 N. Fremont Ave., Idaho Falls, ID 83415}
\affiliation[2]{Massachusetts Institute of Technology, 77 Massachusetts Ave., Cambridge, MA 02139}

%%% Write text for abstract
%%% Most text modifying commands will work in abstract
\begin{abstract}%
Integrated energy systems (IES) are complex heterogeneous architectures that typically encompass power sources, hydrogen electrolyzers, energy storage, and heat exchangers. This integration is achieved through operating control strategy optimization. However, the lack of physical understanding as to how these systems evolve over time introduces uncertainties that hinder reliable application thereof. Techniques that can accommodate such uncertainties are fundamental for ensuring proper operation of these systems. Unfortunately, no unifying methodology exists for accommodating uncertainties in this regard. That being said, adaptive control (AC) is a discipline that may allow for accommodating such uncertainties in real-time. In the present work, we derive an AC formulation for linear systems in which all states are observable, and apply it to the control of a glycol heat exchanger (GHX) in an IES. Based on prior research in which we quantified the uncertainties of the GHX's system dynamics, we introduced an error of 50\% on four terms of the nominal model. In the case where a linear quadratic regulator is used as the nominal control for the reference system, we found that employing AC can reduce the mean absolute error and integral time absolute error by a factor of 30\%--75\%. This reduction is achieved with minimal computing overhead and control infrastructure, thus underscoring the strength of AC. However, the control effort induced is significant, therefore warranting further study in order to estimate its impact on a physical system. To address further challenges, including partially observable and non-linear dynamics, enhancements of the linear formulation are currently being developed.
\end{abstract}



%% Keywords

\end{frontmatter}

%\begin{multicols}{2}
\printnomenclature 

%\nomenclature{\(QoIs\)}{Quantities of interests} 
% --- Metrics
\nomenclature{\(MAE\)}{mean absolute error}
\nomenclature{\(ITAE\)}{integral time absolute error} 
\nomenclature{\(CE\)}{control effort}
\nomenclature{\(INL\)}{Idaho National Laboratory}
% --- Machine Learning
\nomenclature{\(AI\)}{artificial intelligence}
\nomenclature{\(ODE\)}{ordinary differential equation}
\nomenclature{\(SINDyC\)}{Sparse Identification of Nonlinear Dynamic with Control}
\nomenclature{\(MvG-SINDyC\)}{Multivariate Gaussian Fit SINDyC}
\nomenclature{\(GP\)}{Gaussian process}
% --- Control
\nomenclature{\(MRAC\)}{model reference adaptive control}
\nomenclature{\(PID\)}{proportional integral derivative}
\nomenclature{\(AC\)}{adaptive control} 
\nomenclature{\(LQR\)}{linear quadratic regulator}
\nomenclature{\(MPC\)}{model predictive control}
\nomenclature{\(RL\)}{reinforcement learning} 
% --- IES
\nomenclature{\(IES\)}{integrated energy systems}
\nomenclature{\(TES\)}{thermal energy storage}
\nomenclature{\(GHX\)}{glycol heat exchanger}
% --- Misc
\nomenclature{\(NPP\)}{nuclear power plant}
\nomenclature{\(TEDS\)}{Thermal Energy Distributed System}
\nomenclature{$\mu$Rs}{microreactors}
\nomenclature{SMRs}{Small Modular Reactors}
\nomenclature{CPU}{central processing unit}
%\end{multicols}


\newcolumntype{L}[1]{>{\raggedright\arraybackslash}p{#1}}
\renewcommand{\hl}[1]{#1}

%%%%%%%%%%%%%%%%%%%%%%%%%%%%%%%%%%%%%%%%%%%%%%%%
\section{Introduction}
\label{sec:introduction}
%%%%%%%%%%%%%%%%%%%%%%%%%%%%%%%%%%%%%%%%%%%%%%%%
Integrated energy systems (IES), which combine multiple energy sources (e.g., nuclear power), frequently incorporate thermal energy storage (TES) and heat exchangers to maximize energy utilization, manage peak load fluctuations, and address demand-side responses \cite{SEURIN2026111865,seurin2022h2}. These elements work in tandem to improve the overall efficiency and economic benefits of each component within the system. Various sectors of the industry---particularly those involving emerging small modular reactors (SMRs) and microreactors ($\mu$Rs), as well as data centers---stand to benefit from IES. SMRs and nuclear $\mu$Rs cannot compete in the legacy electricity commodity market \cite{Candido24072024}, which is a main reason for the lack of new nuclear builds \cite{buongiorno2019nuclear}. However, IES could increase these reactors' economic value proposition for both grid and non-grid nuclear power plant (NPP) electricity applications \cite{park2025bottom}. The benefit goes both ways, as NPPs can provide baseload, clean, and reliable electricity and heat for hydrogen electrolyzers, industrial applications, or small communities---all of which are key pillars of IES. Enabling proper operation of such systems would lower the barrier to entry for these new reactor types, thereby enabling massive deployment and driving a secure, resilient American energy landscape. This supports the current U.S.~administration's bold ambition to quadruple the nation's nuclear capacity by 2050 \cite{goff2025key}. Moreover, driven by advancements in artificial intelligence (AI), the power demand is expected to surge by more than 160\% by 2030 \cite{seurin2025impact,anderson2024microgrid,aljbour2024powering}, compelling the massive deployment of data centers. IES represent a strategy to power these data centers efficiently, ensuring AI leadership for achieving national security missions and U.S.~economic competitiveness.

However, the complexity of coupling together different assets introduces operational challenges, and classical proportional integral derivative (PID) controllers may be inefficient in this regard \cite{lin2024development,jayaram2025model,das2024design}. Efficient controllers can orchestrate asset cooperation more effectively than can human operators, enhancing both safety and energy utilization and thus generating significant cost savings \cite{dave2023design}. Advances in control methodologies, including model predictive control (MPC) \cite{oncken20224adaptive} and reinforcement learning (RL) \cite{seurin2024assessment,seurin2024multiobjective,TUNKLE2025101090,chen2022deep}, along with access to vast computing resources, offer new pathways for developing advanced controllers capable of performing control actions faster and more efficiently than humans \cite{SEURIN2026111865}. However, the operating environments of these systems can be harsh. The complex operational interconnections between reactors and other resources such as electrolyzers, data centers, district heating, desalination, or co-generation introduce stress due to thermal and mechanical cycling. Additionally, these environments are continuously evolving---radiation exposure, fuel evolution, and mechanical vibrations from the nuclear side for NPPs \cite{abdulraheem2025aload}, and uncertainties and volatility in customer demand from the grid side. For the power production side, intermittent energy production subjected to the vagaries of the weather (e.g., erratic wind speeds) introduce significant uncertainties in forecasting, thus hindering effective decision-making \cite{seurin2025control,lv2019model}. Lastly, equipment aging and wear and tear are a challenge for every single asset, especially within these complex coupled architectures. Consequently, controllers must exhibit robust performance in harsh environments over extended timescales. However, no unifying methodology exists to address these challenges, some of which can be predicted \cite{oncken20224adaptive} despite many uncertainties remaining unknown. MPC is naturally robust to uncertainties via its inherent feedback mechanism \cite{seurin2025control,lv2019model,ji2025model,dong2024model}, and RL exhibits a high level of robustness as well \cite{chen2022deep,TUNKLE2025101090}. However, these advantages are typically demonstrated only in simulations, as the mathematical proofs are non-existent. Thus, a principled approach to dealing with these uncertainties would be of great benefit.

In a recent study \cite{SEURIN2026111865}, we employed uncertainty quantification to tackle these inconsistencies by calibrating a probabilistic Sparse Identification of Nonlinear Dynamics with Control (SINDyC) model called Multivariate Gaussian Fit SINDyC (MvG-SINDyC). Due to the absence of experimental data, we initially generated synthetic data by using Modelica \cite{kral2018modelica}, trained thousands of SINDyC models in order to obtain a distribution of the model’s coefficients, and then fitted a multivariate Gaussian to the distribution. The goal was to encompass all the available experimental data within the posterior probabilistic prediction of MvG-SINDyC, while also considering both uncertainties and the dynamic behaviors of the physics-based models. In \cite{seurin2025control}, we deployed the MvG-SINDyC within a robust MPC framework and used the uncertainty bounds as a penalty for evaluating the constraints. We also found that using a Gaussian process (GP) to correct for model form error was pivotal for ensuring adequate tracking errors within these uncertainties. However, this approach necessitated evaluating the uncertainties in advance and then training a GP. This complex two-level design methodology resulted in a fairly slow MPC controller, especially when the model form error was high, and thus the corresponding uncertainties  \cite{seurin2025control}. This slowness hampers practical tuning of the controller, and several constraints were violated.

Thus, the objective of the present work is to lay the groundwork for developing a self-regulating controller that adapts in real-time to discrepancies (arising from environmental unknown uncertainties) between the expected and the actual behavior of the system. To address this challenge, we propose implementing a robust autonomous controller composed of an outer loop featuring a nominal controller---which can be anything ranging from a PID \cite{jayaram2025model} to a linear quadratic Gaussian  \cite{Vajpayee2021l1adaptive} to RL \cite{annaswamy2023integration}---and an inner loop with AC \cite{narendra2012stable} to guarantee system stability despite the presence of uncertainties.

AC can encompass various types of tools, and while some new applications have burgeoned in recent years in the IES field, AC application remains in its infancy \cite{jayaram2025model,das2024design,deghfel2024new}. The authors of \cite{jayaram2025model} derived a model reference adaptive control (MRAC) controller applicable to hybrid renewable energy systems. MRAC was introduced to improve power flow management in direct current microgrid systems, and efficiently corrected a PID that displayed tracking errors of 30\%--40\%. In \cite{das2024design}, back-stepping MRAC was introduced for microgrid control---including photovoltaics, wind energy, and fuel cells---and, when compared against an adaptive neuro-fuzzy inference system and a fuzzy controller, demonstrated superior stability and speed, as was made noticeable by measuring the quality of power. \cite{deghfel2024new} implemented a two-level maximum power point tracking controller designed for photovoltaic devices with an adaptive neuro-fuzzy inference system as a nominal controller, with genetic and whale optimization algorithms being used to tune the controller parameters. The performance of the two-level maximum power point tracking controller was compared against that of a legacy maximum power point tracking algorithm (namely the incremental conductance method), demonstrating faster convergence times, better tracking efficiency, and improved stability.

While these applications were successful, the assumptions involved can sometimes be opaque and difficult to validate. In practical applications such as IES, it is essential that techniques be interpretable, reliable, and safe. This requirement presents a major challenge for widespread implementation of new AI-based technologies within tightly regulated sectors such as nuclear \cite{hall2024barriers}, thus widening the disparity between AI's potential and its limited adoption in these areas. Therefore, employing interpretable methods for controlling IES is crucial and will advance AI technologies closer toward real-world application. Here, we will derive transparent mathematical assumptions for AC and demonstrate its effectiveness at recovering experimental trajectories for one specific IES (i.e., the Thermal Energy Distributed System (TEDS) at Idaho National Laboratory (INL)) \cite{lin2024uncertainty,frick2020development}, in which our assumptions hold. It is also important that these methods introduce minimal computing overhead and additional infrastructure for control. We thus intend to package our use case and algorithms and release them to the community for purposes of reproducibility and deployment. Lastly, it should be noted that this work is intended to complement a  companion paper \cite{seurin2025control} in which a GP was utilized to correct for model form errors in the IES. To complement the approach, AC could then be applied when the model form errors changed as the operating environment evolved. The contributions of this work can be summarized as follows:
\begin{enumerate}
    \item Derive a rigorous and transparent set of mathematical assumptions for AC and ensure they hold true for the use case, thus ensuring not only robustness but also transparency and reliability.
    \item Apply a methodology with minimal infrastructure and computing overhead (seconds compared to hours in \cite{seurin2025control}), which is conducive to online control.
    \item Achieve first-of-a-kind application of AC to the operation of the INL IES.
%    \item Release of an open-source software for seamless AC that the research community can benefit.
\end{enumerate}

The remaining sections of this paper are organized as follows. Section \ref{sec:adaptivecontrol} defines AC (Section \ref{sec:definition}) and gives the derivation of the specific approach utilized (Section \ref{sec:apracticalalgorithm}). Section \ref{sec:designofexperiments} describes the IES challenge to which we are applying our framework (Section \ref{sec:ghxmodel}), along with the controller and the desired trajectories (Section \ref{sec:controlapproaches}). The results and analysis are showcased in Section \ref{sec:resultsandanalysis}, and concluding remarks and areas of future research are given in Section \ref{sec:conclusion}.
 
%%%%%%%%%%%%%%%%%%%%%%%%%%%%%%%%%%%%%%%%%%%%%%%%
\section{Adaptive Control}
\label{sec:adaptivecontrol}

%%%%%%%%%%%%%%%%%%%%%%%%%%%%%%%%%%%%%%%%%%%%%%%%
\subsection{Definition}
\label{sec:definition}
Let us express a general control dynamic as follows:
\begin{align}
    \dot{x} = f(x,u,t,\theta_x) \\
    y = g(x,u,t,\theta_y)
\end{align}
where $x \in \mathbb{R}^n$ is the state of the system, $y \in \mathbb{R}^p$ represents all the measurable inputs, $y_r \in \mathbb{R}^p$ is a reference tracking signal, $u\in \mathbb{R}^m$ contains the input to the system, and it is often the case that $n >> p > m$. $f$ and $g$ are the system dynamics, all of which may be unknown. That being said, the structures of $f$ and $g$ (order of the system, non-linear relationships, etc.) are often assumed to be known. The system parameter $\theta = (\theta_x,\theta_y) \in \mathbb{R}^l$ is what encapsulates the unknowns, which, as explicated in Section \ref{sec:introduction}, originate from incomplete knowledge of the environment, noises, disturbances, aging, or other unknown phenomena \cite{Vajpayee2021l1adaptive}. The goal of a robust control problem is to design an input $u \in \mathbb{R}^m$ such that an underlying cost function $J(y - y_r,x,u,t,\theta)$ $\forall t>0$ is minimized yet still remains able to handle unexpected and uncertain changes in the system we are trying to control \cite{narendra2012stable}. Thus, the essence of AC lies in introducing an online parameter estimation algorithm (i.e., an \emph{adaptive} component) for $\theta$ and $u$, such that the tracking goals are achieved despite the presence of these uncertain and time-varying systems \cite{annaswamy2023adaptive}.

If we express the tracking error $e(t)$ as $e(t) = y(t) - y_r(t)$, the central goal of the AC is to ensure that $\lim_{t \rightarrow \infty} ||e(t)|| = 0$ while also guaranteeing that the closed-loop system has bounded solutions in real-time. Once this is achieved, the second goal is to ensure learning of the parameter $\theta$ toward a $\theta^{\star}$. This is often referred to as ``perfect learning,'' and is often necessary to avoid undesirable bursts of the tracking error \cite{annaswamy2023adaptive}. Lastly, once these two conditions are satisfied, the cost function can be minimized. This sequence is referred to as the adapt-learn-optimize sequence \cite{annaswamy2023integration}. A typical solution of the AC takes the following form:
\begin{align}
    u := h_u(\theta,\Phi(t),t), \quad
    \dot{\theta} := h_{\theta}(\theta,\Phi(t),t)
\end{align}
where $\Phi(t)$ represents all the data available at time $t$. The functions $h_u(\cdot)$ and $h_{\theta}(\cdot)$ explicitly depend on time, as they are subjected to exogenous signals, including changes in set points and command signals. The difficulty lies in designing $h_u(\cdot)$ and $h_{\theta}(\cdot)$ while ensuring that the overall adaptive system remains stable.

A model reference approach (e.g., MRAC, where a signal $y_r(t)$ is constrained by a reference plant system that $y(t)$ can follow) provides a structure that allows a controller solution $u$ to exist so as to guarantee output tracking. A tractable procedure for determining the structure of $h_u(\cdot)$ and $h_{\theta}(\cdot)$ then relies on a two-step design methodology: the former is determined algebraically, the latter analytically. A large body of literature has been applied to derive the design methodology for various classes of linear and nonlinear dynamical systems \cite{annaswamy2023adaptive}. In Section \ref{sec:apracticalalgorithm}, we specifically derive one for linear systems with nonlinear \emph{matched} dynamics. 
%%%%%%%%%%%%%%%%%%%%%%%%%%%%%%%%%%%%%%%%%%%%%%%%

%%%%%%%%%%%%%%%%%%%%%%%%%%%%%%%%%%%%%%%%%%%%%%%%
\subsection{A practical algorithm for adaptive control}
\label{sec:apracticalalgorithm}
%%%%%%%%%%%%%%%%%%%%%%%%%%%%%%%%%%%%%%%%%%%%%%%%
In the present work, we assume that the state is fully observable, meaning $y = x$ and $n = p$. While cases in which $n \ne p$ can be handled via adaptive observers \cite{annaswamy2023adaptive} or online state estimation techniques such as the extended Kalman filter \cite{abdulraheem2025aload}, we leave that for future research.

If we aim to minimize the cost function $J(x(t)-x_r(t),u,t) = \int_0^{\infty}c(x(t)-x_r(t),u(t))dt$, where $c(\cdot,\cdot)$ is a cost function (e.g., a quadratic regulator \cite{Vajpayee2021l1adaptive}) and the system dynamics are given by $\dot{x} = F(x(t),u(t))$, we can start by linearizing $F$ around an equilibrium point $(x_{eq},u_{eq})$ (set at the origin, for simplicity's sake) such that:

\begin{equation}
\dot{x} = Ax + Bu + f(x,u)\label{eq:referencesystem}
\end{equation}
where $A$ and $B$ are the Jacobian matrices of $F$ with respect to $x$ and $u$, respectively, as evaluated at the equilibrium point. This linearization provides a basis for designing the adaptive controller by simplifying the dynamics to a linear form that can be more easily managed and controlled using established linear control techniques. The uncertainties pertaining to F are factored in $A$, $B$, and $f$. To ensure robust mathematical guarantees, it is always necessary to specify explicit assumptions about the system and verify whether they hold for the plant studied.

\textbf{Assumption 1:} The nonlinearities depend exclusively on x, hence $f(x,u) = f(x)$, which is valid for the case studied (see Section \ref{sec:ghxmodel}). Additionally, $f(x)$ lies within the span of B; therefore, it can be written as $Bf_1(x)$ such that $\dot{x} = Ax + B(x + f_1(x))$. This assumption is the \emph{matching} non-linear hypothesis referred to at the end of Section \ref{sec:definition}.

\textbf{Assumption 2:} Let $A_r,B_r,f_{1,r}$ be the nominal, known values of $A$, $B$, and $f_1$, respectively. We will assume $B = B_r \Lambda$, where $\Lambda$ is symmetric positive definite with $||\Lambda|| \le 1$. This can be interpreted as a loss of control effectiveness (e.g., that induced by malevolent cyberattacks that delay the control actions, and result in damage, or aging of the components. 

\textbf{Assumption 3:} $f_1(x) = \theta_{l,r}\phi(x)$, where $\theta_{l,r}$ is unknown and $\phi(x)$ is a known non-linear mapping, which can arise from modeling error or disturbances. If $\phi(x)$ is unknown, it is common to leverage an approximation from data by fitting a known function form such as an universal approximator (e.g., a radial basis function) \cite{annaswamy2023adaptive}.

\textbf{Assumption 4:} Let $A_h$ be Hurwitz and $\theta_r^{\star}$ such that $A_r + \theta^{\star} B_r = A_h$. We assume the existence of $\theta^{\star}$, such that $A + B\theta^{\star} = A_h$. This is called a matching condition and is ubiquitous in AC \cite{annaswamy2023integration}.

Under these assumptions, the state-space system becomes:
\begin{equation}
    \dot{x} = Ax + B_r\Lambda(u + \Lambda^{-1}\theta_{l,r}\phi(x))
    \label{eq:closedloop}
\end{equation}

And the error model $e(t) = x(t) - x_r(t)$ becomes: 

\begin{equation}
\dot{e} = A_he + B_r\Lambda (u - \theta \Phi(t))    
\end{equation}
where $\theta = [\Lambda^{-1},-\Lambda^{-1}\theta_{l,r},\theta^{\star}]^T$ and $\Phi(t) = [u_{r} - \theta^{\star}_rx_r+f_{1,r}(x_r),\phi(x),x]^T$. As explicated in Section \ref{sec:definition}, an update that would result in $||e(t)|| \rightarrow 0$ must be found. The latter is achieved using an update of the form:
\begin{align}
    u(t) = \hat{\theta}(t)\Phi(t) \\
    \dot{\hat{\theta}} = -\Gamma B_r^TPe\Phi(t)^T
    \label{eq:adaptivecontroller}
\end{align}
where $\Gamma \succ 0$ is positive definite; $P=P^T \in \mathbb{R}^{n\times n} \succ 0$ is positive definite and solves an algebraic Lyapunov equality equation $PA_h + A_h^TP = -Q_{lyap}$, where $Q_{lyap} \succ 0$ is a positive definite matrix; and $\tilde{\theta} = \hat{\theta} - \theta$. The error then becomes:
\begin{equation}
    \dot{e} = A_he + B_r\Lambda \tilde{\theta}\Phi(t)
\end{equation}

Using all the assumptions listed, it can be shown that the closed-loop system specified by the reference system \ref{eq:referencesystem}, the plant in Equation \ref{eq:closedloop}, and the adaptive controller \ref{eq:adaptivecontroller} all lead to bounded solutions for $x(t)$, $\hat{\theta}(t)$, and $e(t)$---no matter the initial conditions $x(0)$, $\hat{\theta}(0)$, and $e(0)$. Moreover, if $\Phi(t)$ is bounded $\forall t>0$, then it can be found through a Lyapunov stability analysis in which $\lim_{t \rightarrow \infty} ||e(t)|| = 0$. Boundedness can be achieved by appropriately selecting a controller capable of maintaining the boundedness of the reference system states. This is expressed as $\forall m_1 > 0$, $\exists m_2$ such that if $||x(0)|| \le m_2$, then $||x(t)|| \le m_1$ $\forall t > 0$ \cite{annaswamy2023integration}.  
\\
\textbf{Proof:} Consider the following Lyapunov candidate function: $V(e,\tilde{\theta}) = e^TPe + Tr(\tilde{\theta}^T(\Lambda^TS)\tilde{\theta})$, where $S = \Gamma^{-1}$---with the symmetric part of $(\Lambda S)$ positive definite. It is clear that $V(e,\tilde{\theta}) > 0$, except for $V(0,0) = 0$. Therefore, $V(e,\theta)$ is lower bounded. Solving for $\dot{V}(e,\tilde{\theta}) \le 0$ gives us:
\[
    \dot{V}(e,\tilde{\theta}) = \nabla V\dot{e} = 2e^TP\dot{e} + 2\tilde{\theta}^T(\Lambda^TS) \dot{\tilde{\theta}}
\]
\[    \dot{V}(e,\tilde{\theta}) = \nabla V\dot{e} = 2e^TP(A_he + B\Lambda(u+\Lambda^{-1}\theta_{l,r})) +  2\tilde{\theta}^T(\Lambda^TS) \dot{\tilde{\theta}}
\]
\[    \dot{V}(e,\tilde{\theta}) = \nabla V\dot{e} = -e^TQ_{lyap}e + 2e^TPB\Lambda(u -\theta\Phi(t)) +  2\tilde{\theta}^T(\Lambda^TS) \dot{\tilde{\theta}}
\]

With our choices of updates for $u$ and $\dot{\tilde{\theta}}$, we obtain $\dot{V}(e,\tilde{\theta}) = -e^TQ_{lyap}e \le 0$. Hence, $\dot{V}$ is negative semi-definite and thus $e$ and $\tilde{\theta}$ are uniformly bounded. If $\Phi(t)$ is bounded, then so is $\dot{e}$, and $e\in \mathcal{L}_2$. That being the case, $\ddot{V} = -2e^TQ_{lyap}\dot{e}$ is bounded and therefore $\dot{V}$ is uniformly continuous. Per Barbalat's lemma (see \ref{appendix:convergenceofthetrackingerror}), $\lim_{t \rightarrow \infty} \dot{V}(e,\tilde{\theta}) = 0$; therefore, $lim_{t \rightarrow \infty} ||e(t)|| = 0$.

For the glycol heat exchanger (GHX) applications (See Section \ref{sec:ghxmodel}) and closely related nuclear subsystems, the assumptions can be easily verified. First, the state is fully observable in our use case ($y=x$) and the identified linear model is fully controllable with two actuators. Moreover, the GHX reduction used in this paper directly enables the construction of $\theta^\star = B^{-1}(A - A_h)$ required by Assumption~4. Boundedness of $\Phi(t)$ follows from bounded reference trajectories and the stabilizing LQR that defines $A_h$. In addition, the assumed matched nonlinearity $f_1(x) = \theta_{l,r}\,\phi(x)$ (Assumption~3) can be used in angle-actuated components prevalent in IES and nuclear systems. For example, the reactivity worth of rotating control drums varies with drum angle and is commonly represented by a cosine-based model derived from first-order perturbation theory, making $\sin\theta$ and $\cos\theta$ natural basis functions for~$\phi(x)$ \cite{TUNKLE2025101090}. On the balance-of-plant side, quarter-turn butterfly valves exhibit smooth, strongly angle-dependent $C_v(\alpha)$ characteristics, which can be well captured by $\sin\theta$ and $\cos\theta$ over normal operating ranges. Collectively, these examples make the $\sin/\cos$ choice both interpretable and easy to validate for our target systems. %However, development of the AC algorithm also hinges on the existence of a $\theta^{\star}$ that satisfies the matching condition of Assumption 4. Its existence can sometimes be demonstrated by examining the expected uncertainties that arise from the target parameters, then identifying the pole placement strategy $u = \theta^{\star} x$ that could result in the Hurwitz matrix of interest $A_h$.

\subsection{Extension for matched unknown disturbances}
\label{sec:extension}
In the previous section, we detailed how to recover an exact tracking error when the shape of the modeling inadequacy is known. However, it is also possible to account for unknown bounded disturbances and derive a guarantee that the error will be bounded. These disturbances can occur due to the stochastic nature of neutron interactions in the NPPs, fluctuation in temperature and pressure, and vibration of internal parts \cite{Vajpayee2021l1adaptive}. The matched disturbance was injected into the control rod speed as a chirp signal in \cite{Vajpayee2021l1adaptive}, such that:

\begin{equation}
    x(t) = Ax(t) + B_r\Lambda(u(t) + \Lambda^{-1}\Theta_{l,r}\phi(x) + \Lambda^{-1} d(t))
    \label{eq:matcheddisturbance}
\end{equation}
 where $||d(t)|| \le d_{max}$. 
 
 \textbf{Proof:} In this instance, a general derivation is proposed in connection with the $\sigma$-adaption law \cite{park2020model}, with the update being of the following form:
\begin{equation}
    \dot{\hat{\theta}} = -\gamma B^TPe\Phi^T - \sigma\hat{\theta}
\end{equation}
where, if $\sigma$ is a positive constant, then $e$ and $\tilde{\theta}$ are bounded $\forall t \ge 0$. If we consider $V = \frac{1}{2}e^TPe + \frac{1}{2}Tr(\tilde{\theta}^T(\Lambda S)\tilde{\theta})$ and $D = B d$, we obtain:
\begin{equation*}
    \dot{V} = -e^TQe + 2e^TPBd - \tilde{\theta}^T\sigma \tilde{\theta} - \tilde{\theta}^T\sigma \theta^{\star} \le -\lambda_{min}(Q)||e||^2 + 2 ||e|| ||PD|| - \sigma ||\tilde{\theta}||^2 - \sigma ||\tilde{\theta}||||\theta^{\star}||
\end{equation*}
\begin{equation*}
    = - \lambda_{min}(Q)(||e|| - \frac{||PD||}{\lambda_{min}(Q)})^2 + \frac{||PD||^2}{\lambda_{min}(Q)} - \sigma(||\tilde{\theta}||^2 + 2||\tilde{\theta}||\frac{||\theta^{\star}||}{2} + \frac{||\theta^{\star}||^2}{4} - \frac{||\theta^{\star}||^2}{4}) 
\end{equation*}
\begin{equation*}
    = - \lambda_{min}(Q)(||e|| - \frac{||PD||}{\lambda_{min}(Q)})^2 + \frac{||PD||^2}{\lambda_{min}(Q)} + \sigma \frac{||\tilde{\theta}||^2}{4} - \sigma(||\tilde{\theta}|| + \frac{||\theta^{\star}||}{2})^2
\end{equation*}
Let $\Psi = \{(e,\tilde{\theta}); \quad \lambda_{min}(Q)(||e|| - \frac{||PD||}{\lambda_{min}(Q)})^2 +  \sigma(||\tilde{\theta}|| + \frac{||\theta^{\star}||}{2})^2 \le \frac{||PD||^2}{\lambda_{min}(Q)} + \sigma \frac{||\tilde{\theta}||^2}{4}\}$. The origin is $(0,0)\in \Psi$, and because $d$ is bounded, $D$ is also, and therefore $\Psi$ is a bounded neighborhood of the origin. Moreover, $\forall (e,\tilde{\theta}) \in \Psi^c$, where $\Psi^c$ is the complementary set of $\Psi$, $V(e,\tilde{\theta}) > 0$ and $\dot{V}(e,\tilde{\theta}) \le 0$. On top of that, $lim_{(e,\tilde{\theta}) \rightarrow \infty}V(e,\tilde{\theta}) = \infty$. Per the Lagrange stability theorem \cite{lasalle1960some}, $e$ and $\tilde{\theta}$ are bounded $\forall t \ge 0$. Moreover, every $e$ and $\tilde{\theta}$ that starts in $\Psi^c$ will enter $\Psi$ and remain there thereafter; therefore, $e$ and $\tilde{\theta}$ are bounded, and also $e = O(max(d_{max},||\theta^{\star}||))$.  

%%%%%%%%%%%%%%%%%%%%%%%%%%%%%%%%%%%%%%%%%%%%%%%%%
%\subsection{Recent related works}
%\label{sec:relatedworks}
%%%%%%%%%%%%%%%%%%%%%%%%%%%%%%%%%%%%%%%%%%%%%%%%%
%In this Section, we will summarize recent works in the area of AC for nuclear applications and emphasize how our work differentiate form them: 

%The authors in \cite{oncken20224adaptive} derived an Adaptive MPC that modify the systems' dynamic based on the detection of  Heat Pipe (HP) failures in a HP microreactor with a decision tree. They found that they could track a reference trajectory reasonably well, while satisfying temperature constraints. However, the system's dynamic in normal and degraded states must be known in advance, and the performance of the overall workflow hinges upon the detector, which cannot be predicted with mathematical robustness.

%In \cite{abdulraheem2025aload}, the authors leveraged a super-twisting SMC for load-follow of the Holos-Quad microreactor, a high temperature gas-cooled graphite mpoderated reactor. They compared it against PID and nonlinear MPC on a nonlinear point kinetic equation with thermal modelling and xenon and iodine feedbacks under parameter variations and load changes. They found that over long period nonlinear MPC surpasses both PID and SMC, but to the cost of higher computational resources. 
The updates in the control law are applied to the estimated states of the system, where they assumed a known dynamic. Over time, this constant schedule may be undesirable in the event of an unpredictable---or unstable---response. While it accounts for disturbances in the system, our approach proposes a solution that deals with time-varying parameters.

%\cite{Adaptive neural network sliding mode controller design for load following
%of nuclear power plant} mentions it.

%\cite{Model Reference Robust Adaptive Control of Control Element Drive
%Mechanism in a Nuclear Power Plant} mentions it.

%A closer approach to our work incudes  \cite{Vajpayee2021l1adaptive}, where the authors apply an $L_1$-adaptive (an advanced model reference adaptive control (MRAC)) state feedback controller using a Linear Quadratic Gaussian (LQG) as a nominal controller for Pressurized Water Reactor (PWR) control for various sets of tracking trajectories, including regular and emergency transients. Subsystems included power, steam generator, pressurizer, and turbine speed controls were considered, with perturbations in coefficients, matched, and unmatched disturbances. They found that their approach resulted in uniform and smoother transient with better tracking error, and lower control effort, which is beneficial for reducing damage on equipment while guaranteeing high performance. %Here, we propose to apply an instance of MRAC to a very challenging task of BWR load-follow under density-wave oscillations where uncertainty of the system result in very large change in state's stability \cite{DIROCCO2020107749}. 

%Moreover, we are very clear about our assumption to ensure that both empirically but also theoretically it works.

%%%%%%%%%%%%%%%%%%%%%%%%%%%%%%%%%%%%%%%%%%%%%%%%
\section{Design of Experiments}
\label{sec:designofexperiments}
%%%%%%%%%%%%%%%%%%%%%%%%%%%%%%%%%%%%%%%%%%%%%%%%

%%%%%%%%%%%%%%%%%%%%%%%%%%%%%%%%%%%%%%%%%%%%%%%%
\subsection{GHX model}
\label{sec:ghxmodel}
%%%%%%%%%%%%%%%%%%%%%%%%%%%%%%%%%%%%%%%%%%%%%%%%
The TEDS testbed at INL evaluates the performance of IES, with the TES and the ethylene-glycol-to-Therminol-66 GHX having been modeled using SINDyC \cite{lin2024development}. An MPC-based controller was then developed to assess advanced control efficacy within this system. Nevertheless, discrepancies between the model data and the physical experiments hindered practical deployment.

In a previous study \cite{SEURIN2026111865}, we developed a SINDyC \cite{brunton2016sparse} ensemble model for uncertainty quantification of the GHX based on a real experimental test set that had been augmented with synthetic data from a Modelica model. Subsequently, we attempted to apply MPC \cite{seurin2025control}. The SINDyC model formulation utilizes an affine library $1, x$, resulting in an ordinary differential equation of the form:

\begin{equation}
\frac{dx}{dt} = A_{ghx}x + B_{ghx}u + D_{ghx}
\label{eq:ghxall}
\end{equation}
where $A_{ghx}\in \mathbb{R}^{2\times 2}$ and $B_{ghx}\in \mathbb{R}^{2\times 4}$ and $D_{ghx}\in \mathbb{R}^{2\times 1}$. This then gives us:

$x = \left(\begin{array}{c} \dot{m}_{ghx,bypass} \\ Q_{ghx} \end{array} \right)$ and $u =  \left(\begin{array}{c} PV_{006} \\ \dot{m}_{pump,out} \\ T_{pump,in} \\ T_{heater,out} \end{array} \right)$

\begin{figure}[htp!]
    \centering
\includegraphics[width=0.9\linewidth]{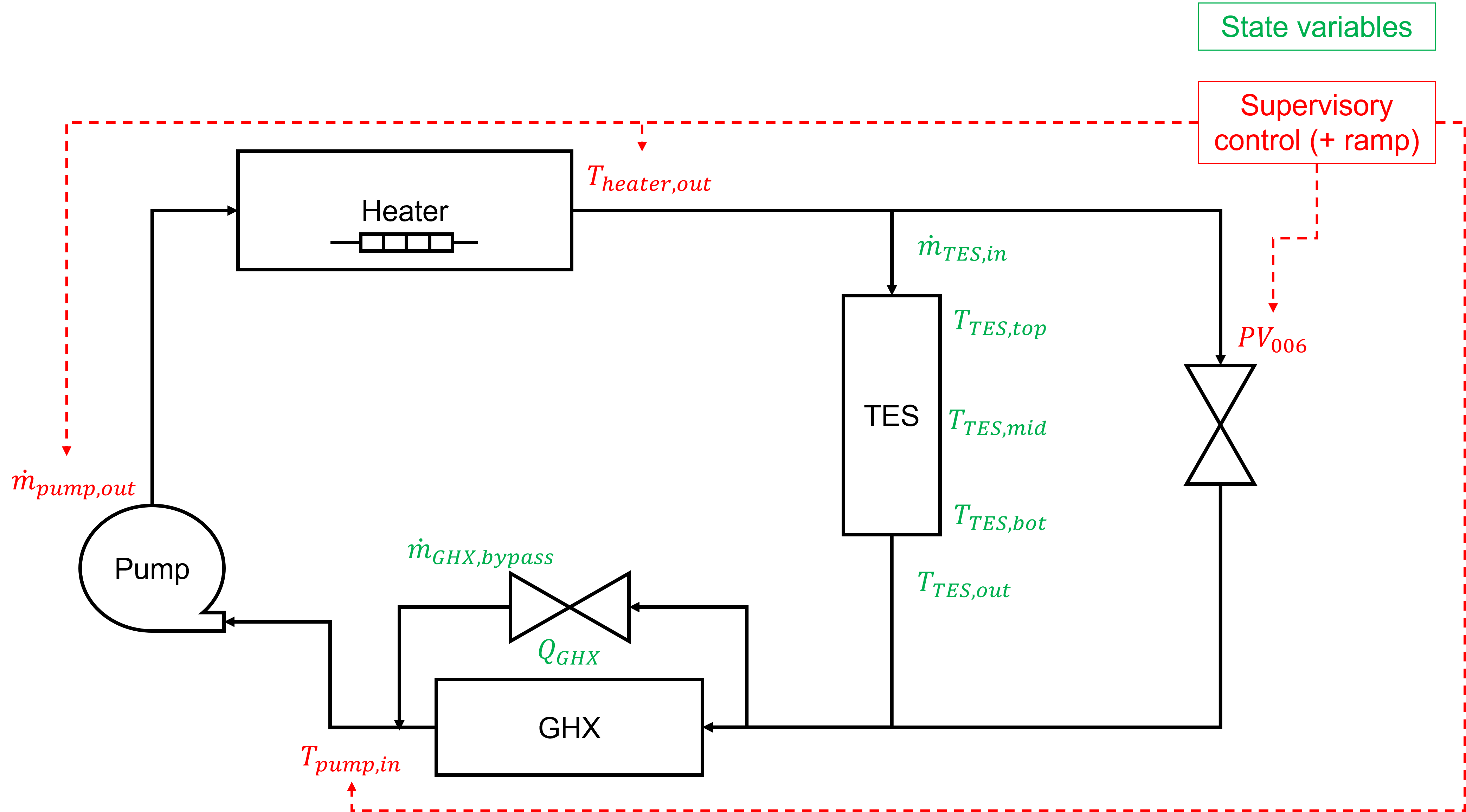}
    \caption{Simplified TEDS and control systems. Though a model of TES was also developed, the model form error was too large, necessitating that a GP correction be added, which is why we have omitted it from this study.}
    \label{fig:tedsandcontrol}
\end{figure}

A schematic of TEDS is given in Figure \ref{fig:tedsandcontrol}.

\begin{figure}[htp!]
      \centering      
      \includegraphics[width=0.9\linewidth]{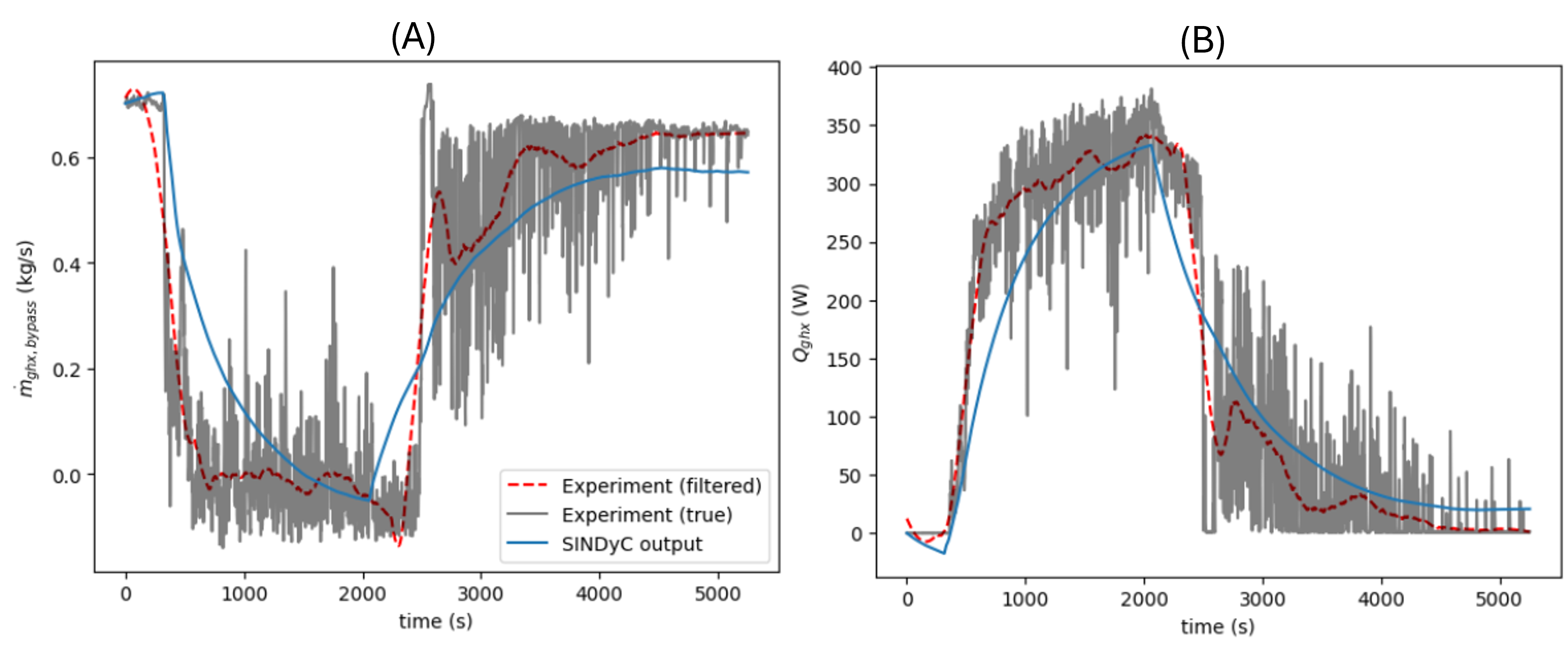}
      \caption{Performance of the SINDyC model on the GHX experiment for (A) $\dot{m}_{ghx,bypass}$ and (B) $Q_{ghx}$. Five hundred equally spaced data points and a polynomial of order 2 were used to obtain the filtered experimental trajectories (red dashed curves).}
      \label{fig:performanceofghx}
  \end{figure}

Nevertheless, as alluded to in Section \ref{sec:definition}, often $p > m$; however, in this case, $p$ is equal to 2 and $m$ to 4. Therefore, this system is over-parameterized. The rationale behind this is that we originally wanted to couple the TES and GHX within a single model. In this instance, we ended up with $p = 7 > m = 4$, which did not work correctly. As a result, here we fit a new SINDyC model based on the two actuators that have the greatest impact on the GHX: $PV_{006}$ and $\dot{m}{pump,out}$, and $u =  \left(\begin{array}{c} PV_{006} \\ \dot{m}_{pump,out} \end{array} \right)$ and $A_{ghx}\in \mathbb{R}^{2\times 2}$ and $B_{ghx}\in \mathbb{R}^{2\times 2}$ and $D_{ghx}\in \mathbb{R}^{2\times 1}$. The data used for training were obtained via filtering the experiments with a Savitzky-Golay filter \cite{SEURIN2026111865}. Figure \ref{fig:performanceofghx}(A) and (B) showcases the new SINDyC model's performance for $\dot{m}{ghx,bypass}$ and $Q_{ghx}$, respectively. The original noisy dataset from the experiment is reflected in the black-shaded curves, the dashed red curves represent the filtered experiment, and the blue curves represent the SINDyC output. While the granularity of the dashed red curves is not completely recovered, the results were deemed sufficient for the purpose of the study. Moreover, we will see in Section \ref{sec:application} that a set of actuators can be found to recover a better match. The matrices composing the dynamics are as follows:
$A_{ghx} = \left[\begin{array}{cc}
-1.27006037e-03  & 0.00000000e+00 \\
 -1.67511974e+00 & -4.89615042e-03
\end{array}\right]$, $B_{ghx} = \left[ \begin{array}{cc}
   -0.00083076 & 0.00462962 \\
 0.51405729 & 0.57604899
\end{array}\right]$, and $D_{ghx} = \left[\begin{array}{c}
     -0.0022987 \\ 0.68611759  
\end{array}\right]$.

Lastly, we also verified that the rank of the rows of the matrix $C_{n-1} = \left[\begin{array}{ccccc} B & AB & A^2B & ... & A^{n -1}B \end{array}\right]$ \cite{seurin2025control} is equal to 2, thus ensuring controllability. \\

With $C_{n-1}=\left[\begin{array}{cccc}
     -8.30757614e-04 & 4.62962252e-03 & 1.05511232e-06 & -5.87990009e-06 \\
  5.14057288e-01 & 5.76048993e-01 & -1.12528333e-03 &-1.05755946e-02
\end{array}\right]$, it is the case that the singular values of $C_{n-1}$ are equal to 0.77214562 and 0.00370246, hence the rank is 2. The condition number is 209, which is reasonable and thus the system was deemed also practically controllable.

%%%%%%%%%%%%%%%%%%%%%%%%%%%%%%%%%%%%%%%%%%%%%%%%
\subsection{Control use case}
\label{sec:controlapproaches}
%%%%%%%%%%%%%%%%%%%%%%%%%%%%%%%%%%%%%%%%%%%%%%%%
In this work, we utilized a LQR to track a reference trajectory $x_{ref}(t)$ and the desired actuators $u_{ref}(t)$. Briefly, an LQR is a controller solution that minimizes $J(x,u) = \int_{0}^{+\infty} (x^TQx + u^TRu) dt$, where $Q$ and $R$ are symmetric-positive-definite cost matrices. At each timestep $u(t) = -R^{-1}B^TP(x(t) - x_{ref}(t))+u_{ref}$, where P is the solution of the Riccati equation \ref{eq:riccatiequation}, which can be solved numerically, $u(t)$ is the optimal controller in the feedback control problem.
\begin{equation}
    A^TP + PA  - PBR^{-1}B^TP + Q = 0
\label{eq:riccatiequation}
\end{equation}
$A$ and $B$, as described in Section \ref{sec:ghxmodel}, are equal to $A_{ghx}$ and $B_{ghx}$, respectively. The trajectory to follow is that of the physical TEDS experiment, as described in \cite{SEURIN2026111865} or in the dashed red curve in Figure \ref{fig:performanceofghx}. For testing the method, we verified in which parameter from $A_{ghx}$, $B_{ghx}$, and $D_{ghx}$ the uncertainty was substantial. Looking at the coefficient's distribution from MvG-SINDyC \cite{SEURIN2026111865}, we found that $A_{ghx}[1,0]$, $B_{ghx}[1,0]$, $B_{ghx}[1,1]$, and $D_{ghx}[1]$ concentrate most of the uncertainties. To test the approach on one more use case, we introduced a matched nonlinearity of the form $f_1(x)=\theta_{l,r} sin(x)\in \mathbb{R}^{2 \times 1}$, where $\theta_{l,r} = 1.5$. Since the model is linear, Assumptions 1--3, listed in Section \ref{sec:apracticalalgorithm}, all hold. the GHX reduction used in this paper directly enables the construction of $\theta^{\star} = (B^TB)^{-1}B^T(A - Ah)$ (or $B^{-1}(A - A_h)$, since $B$ is invertible) required by Assumption~4, where $A_h$ is a Hurwitz matrix using the LQR gain $K=R^{-1}B^TP$, such that $A_h = A_r-B_rK$. 

%%%%%%%%%%%%%%%%%%%%%%%%%%%%%%%%%%%%%%%%%%%%%%%%
\section{Results and Analysis}
\label{sec:resultsandanalysis}
%%%%%%%%%%%%%%%%%%%%%%%%%%%%%%%%%%%%%%%%%%%%%%%%

%%%%%%%%%%%%%%%%%%%%%%%%%%%%%%%%%%%%%%%%%%%%%%%%
\subsection{Metrics utilized to test the performance of the controller}
\label{sec:metricsutilized}
%%%%%%%%%%%%%%%%%%%%%%%%%%%%%%%%%%%%%%%%%%%%%%%%
In this work, we focus on tracking error recovery; therefore, we decided to zoom in on two regulation metrics: mean absolute error (MAE) and integral time absolute error (ITAE):
\begin{equation}
    MAE = \frac{1}{N}\sum_{i=1}^N |e(i)|,   \quad ITAE = \int_{t=0}^T t \times |e(t)|dt
\end{equation}%\quad CAE = \int_{0}^T |e(t)|dt \sim \frac{T}{N}MAE
where $N$ is equal to the number of datapoints (here, 5,251). To measure AC's impact on the actuators, we also introduced the control effort (CE):
\begin{equation}
    CE = \int_{t=0}^T||u(t)||_{l}dt
\end{equation}%, \quad TVI = \sum_{i=1}^N |u_{i+1} - u_i|
where $||\cdot||_{l,1}$ is the $L_{p,1}$ norm, with $l$ equal to 2 in \cite{abdulraheem2025aload} and $1$ in \cite{TUNKLE2025101090}. This will indicate the increased penalty on the control actions as a result of unknown perturbations.
%For the servoregulation objective, the Control Effort (CE) and the Total Variation of Input (TVI) are appropriate
%\begin{equation}
%    CE = \int_{t=0}^T||u(t)||_{l,1}dt, \quad TVI = \sum_{i=1}^N |u_{i+1} - u_i|
%\end{equation}
%where $||\cdot||_{l,1}$ is the $L_{p,1}$ norm with $l$ equal to 2 in \cite{abdulraheem2025aload} and $1$ in \cite{TUNKLE2025101090}. This will give an indication of the loss of performance due to perturbation.
%%%%%%%%%%%%%%%%%%%%%%%%%%%%%%%%%%%%%%%%%%%%%%%%
\subsection{Application of AC to the GHX model}
\label{sec:application}
%%%%%%%%%%%%%%%%%%%%%%%%%%%%%%%%%%%%%%%%%%%%%%%%
To demonstrate the viability of the approach, we study two cases: one with only uncertainty in $A_{ghx}$, $B_{ghx}$, and $D_{ghx}$, and one with matched disturbances $f_1(x)$. For the former, we multiply $D_{ghx}[1]$ by 1.5 and divide the uncertain terms in $A_{ghx}$ and $B_{ghx}$ by that same value---hence corresponding to an uncertainty level of 50\%. For $f_1(x)$, we assume $f_1(x) = \theta_{l,r}sin(x)$, where $\theta_{l,r} = \left[\begin{array}{cc}
   1.0  & 0.0\\
     0.0  & 1.0 \\
\end{array}\right]$. The choice of models for $f_1(x)$ is a parsimonious way to represent angle-driven nonlinearities present in GHX actuation. In practice, the bypass butterfly valve and/or upstream dampers follow smooth stroke-to-capacity curves in which effective area---and thus heat transfer and flow---change nonlinearly with blade/disc angle, whose dependencies can be well captured by $\sin\alpha$ and $\sin 2\alpha$ functions in the normal operating domain. The same functional form can also be used in the nuclear/IES contexts highlighted in \ref{sec:apracticalalgorithm} for control drums in microreactors with cosine-based worth models. Note that the uncertainty in $D_{ghx}$ can either be disregarded or explicitly dealt with with Assumption 3:
\begin{itemize}
\item For the former, the uncertainty in $D_{ghx}$, $\tilde{D}_{ghx}$ can be seen as a matched unknown disturbance. Indeed, it exists $d\in \mathbb{R}^{2\times 1}$ such that $B_rd = \tilde{D}_{ghx}$, $d = B_r^{-1}\tilde{D}_{ghx}$ and Equation \ref{eq:ghxall} can be written in the form of Equation \ref{eq:matcheddisturbance}. We demonstrated in Section \ref{sec:extension} that the tracking error will be bounded. 
\item To deal with the uncertainties in $D_{ghx}$ explicitly, we can write $f_1(x)$ as $\theta_{l,r}\sin(x)+B^{-1}_{ghx}\tilde{D}_{ghx}\phi_D(x)$ where $\phi_D(x) = \left[\begin{array}{cc}
   1.0  & 0.0\\
     0.0  & 1.0 \\
\end{array}\right]$, which can be dealt with with an estimator of $\theta_{l,r}$ and $B^{-1}_{ghx}\tilde{D}_{ghx}$ with Assumption~3. 
\end{itemize}
In this work, we will also compare both instances. 

\begin{figure}[htp!]
    \centering  
    \includegraphics[width=1.0\linewidth]{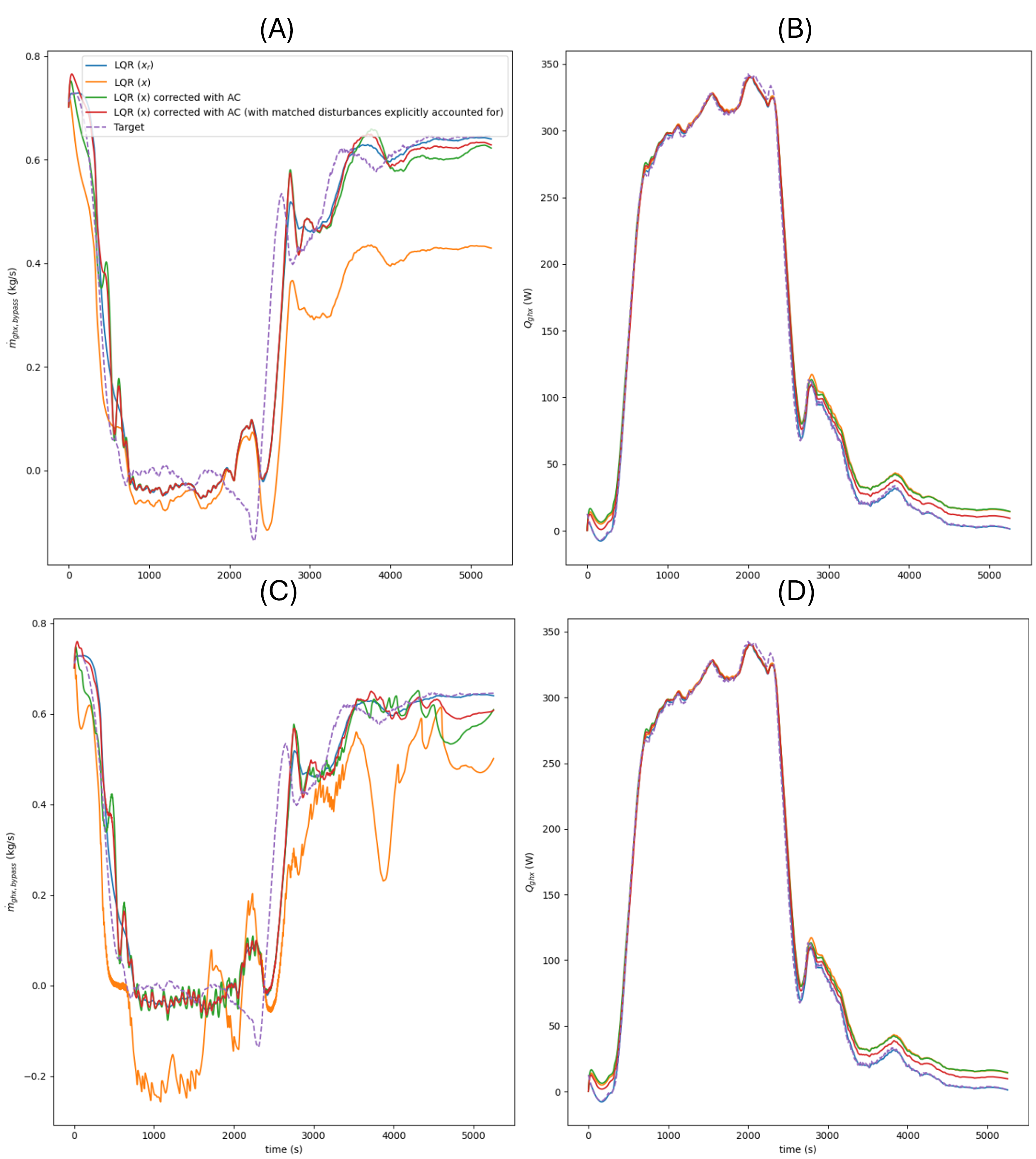}
    \caption{Comparison of the LQR controllers in the presence of uncertainties, both with and without correction from AC: $\theta_{l,r} = 0$ ((A) and (B)) and $\theta_{l,r} = 1.5$ ((C) and (D)).}
    \label{fig:ghxall}
\end{figure}

The hyperparameters for the LQR and AC are summarized in Table \ref{tab:hyperparametersforlqrandac}, and were found through manual tuning. The initial estimate of the parameters are given in Table \ref{tab:initialvalueforparameters}, which must be estimated in practice but given here as indicators to test the performance of the AC. For brevity, we omit both aspects of the study. The $\sigma$ in the $\sigma$-adaption scheme was set to 0, as it primarily helps with stability and was found to be unnecessary here. The results of controlling for the GHX experimental trajectories (described in Section \ref{sec:controlapproaches}) with the LQR, both with and without perturbation, as well as the correction with AC with and without accounting explicitly for the uncertainties in $D_{ghx}$, are showcased in Figure \ref{fig:ghxall}.

The dashed purple lines represent the targets for each state. The blue curves are the trajectories that resulted from applying LQR to the unperturbed (known) dynamics. The orange curves are the trajectories that resulted from applying the LQR gain ($K$) to the perturbed dynamics, and the green and red curves curves are the trajectories that resulted from the latter in tandem with an AC correction without and with accounting for the uncertainties in $D_{ghx}$ explicitly. First, the blue curves align very well with the dashed purple lines, while the orange curves stray away from them, resulting in an MAE of 0.0484 and 2.0617 and 0.147 and 8.885
 for the case without matched disturbances, respectively, for $\dot{m}_{ghx,bypass}$ and $Q_{ghx}$, respectively. For the scenario with disturbances we obtain 0.1466 and 12.455 for $\dot{m}_{ghx,bypass}$ and $Q_{ghx}$, respectively. Along the trajectories, the green and red curves align significantly better with the blue curves than the orange ones. The red curves is always closer, with a slight loss of performance beyond 4,000 seconds. 
 
 This is supported by Tables \ref{tab:maefordifferentepsilonmdot} and \ref{tab:maefordifferentepsilonqghx}, for $\dot{m}_{ghx,bypass}$ and $Q_{ghx}$, respectively. The Tables show that the MAE and ITAE decreased with AC by up to 63\% and 75\%, respectively for $\dot{m}_{ghx,bypass}$ and 50\% to 55\% for the $Q_{ghx}$. The MAE and ITAE were calculated by comparing the experimental trajectories and the trajectories obtained by applying the LQR gain to the perturbed system, LQR $x$, both with and without AC, divide by the MAE and ITAE obtained by comparing the experiment with the resulting trajectories obtained by applying LQR to the reference system, LQR $x_r$, which AC is supposed to help track. Moreover,  the MAE and ITAE obtained from explicitly account for the uncertainty $\tilde{D}_{ghx}$ is systematically lower than that of not accounting for it, where the differences for $Q_{ghx}$ can be rather substantial (about 33\% for both MAE and ITEA). While the error has been proven to be bounded (see Section \ref{sec:extension}), it supports that accouting for the error might be necessary.
 
 The last row of Table \ref{tab:maefordifferentepsilonmdot} compares the computing times for generating the full trajectories, both with and without AC. While AC is about 3 times slower, the worst total computing time (obtained for LQR ($x$) + AC and accounting explicitly for the uncertainties in $D_{ghx}$) is only 0.0011 central processing unit (CPU) seconds per physical second, which is reasonable. Essentially AC does not add any computing overhead, which is very conducive to real time control. Additionally, the results were obtained through online adaptation, without any assumptions on the magnitude of the uncertainties while prescribing the control $u$. These outcomes confirm the efficacy and viability of the AC approach in regard to self-regulating challenging IES. This contrasts \cite{seurin2025control} where we first quantify the uncertainties and deploy a robust MPC model to penalize the magnitude of the worst state with the 95\% confidence interval from these uncertainties (also known as chance constraints). Then, the computing overhead amounted to hours not milliseconds. However, for AC to work the simulation-to-real (i.e., uncertainties) gap must be ``small enough'' \cite{annaswamy2023adaptive}, while large constraints can be accommodated with chance constraints.

\begin{table}[htp!]
    \centering
\caption{Hyperparameters for the LQR and AC. A similar gain $K$ for the LQR and the AC correction (i.e., $\theta^{\star}_{r}$ in Assumption 4) is assumed, as obtained from $Q$ and $R$.}
    \begin{tabular}{c|c|c|c}
    \hline
        $Q$ & 
        $R$ & 
       $Q_{lyap}$  &  
        $\Gamma$ \\
\hline
$10$ & $1000$ & $1e-6$ & $1e-4$ 
    \end{tabular}
\label{tab:hyperparametersforlqrandac}
\end{table}

\begin{table}[htp!]
    \centering
\caption{Initial value for the unknown parameters.}
    \begin{tabular}{c|c|c|c}
    \hline
        $\Lambda_0$ &
        $\theta^{\star}_0$ & $(\theta_{l,r})_0$&
        $(\theta_{l,r})_0^D$\\
\hline
$0.80\times\Lambda$ & $\theta^{\star}_{r}$  &   $0.75\times\theta_{l,r}$   &   $0.75\times\theta_{l,r}^D$  
    \end{tabular}
\label{tab:initialvalueforparameters}
\end{table}

\begin{table}[htp!]
    \centering
      \caption{Normalized MAE and ITAE for $\dot{m}_{ghx,bypass}$ and (CPU) computing time to obtain the whole trajectory  both without (left) and with matched disturbances (right).}
    \begin{tabular}{c|c|c|c|c}
    \hline
         &  LQR ($x$) & LQR ($x$) + AC & LQR ($x$) + AC (explicit account for $\tilde{D}_{ghx}$) & relative difference\\
         \hline
     MAE & 3.0326/3.0313
 & 1.208/1.311 & 1.134/1.194& 63\%/ 61\%\\
 ITAE & 4.737/3.858 & 1.421/1.641 & 1.202/1.362 & 75\%/65\%\\
 Computing time (s) & 1.1/1.4 & 3.9/4.8 & 4.7/6.0 & -327\%/-329\% \\
    \end{tabular}
\label{tab:maefordifferentepsilonmdot}
\end{table}

\begin{figure}[htp!]
    \centering
    \includegraphics[width=0.9\linewidth]{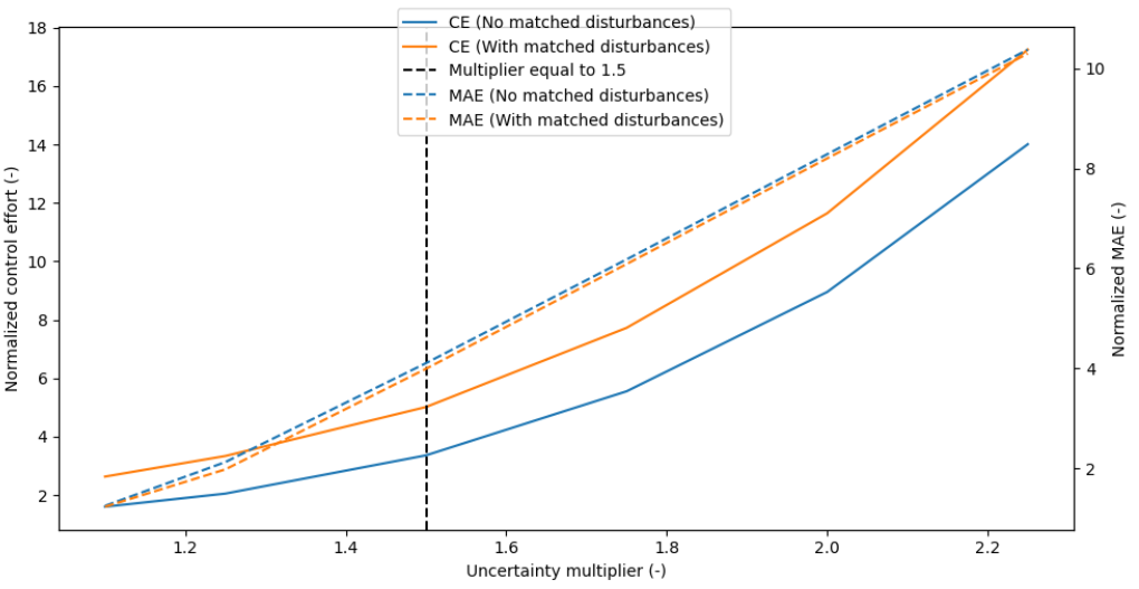}
    \caption{CE and MAE as a function of the multiplier for the case without matched disturbances (blue) and that with matched disturbances (orange). The MAEs and CEs are normalized to the MAE and CE of the nominal case (i.e., LQR applied to the reference system), respectively. Dashed lines are added to help visualize the MAE and CE for the case studied in Figure \ref{fig:ghxall}.}
    \label{fig:normalizedcontroleffort}
\end{figure}

\begin{table}[htp!]
    \centering
      \caption{Normalized MAE and ITAE for $Q_{ghx}$  both without (left) and with matched disturbances (right).}
    \begin{tabular}{c|c|c|c|c}
    \hline
         &  LQR ($x$) & LQR ($x$) + AC & LQR ($x$) + AC (explicit account for $\tilde{D}_{ghx}$) & relative difference\\
         \hline
     MAE & 4.309/6.0409 & 4.239/4.525
 &  2.915/3.0178 & 32\%/50\%\\
 ITAE & 6.100/8.788 & 5.857/6.254& 3.844/3.989 & 37\%/55\%\\
    \end{tabular}
\label{tab:maefordifferentepsilonqghx}
\end{table}
Nevertheless, the tracking error does not tell the whole story. While we can fairly well recover the trajectory with AC, this comes at the cost of increased CEs. For the final aspect of this study, we will examine the impact of uncertainty on the CE, via the multiplier. Here, the uncertainty $\tilde{D}_{ghx}$ is dealt with explicitly. The results are showcased in Figure \ref{fig:normalizedcontroleffort}.
While the MAE for both without and with matched disturbances are very close, the CEs are not, where the CE for the latter is about 1.5 times as that of the former. On top of that, the CE increases sub-linearly at up to a multiplier value of 1.8 with an MAE about 3.5 times that of the nominal case, then begins to follow a closer-to-exponential pattern, suggesting that an uncertainty of 80\% is probably the upper limit. Therefore, while we demonstrated that AC is a powerful tool for recovering the trajectory from an LQR applied to a nominal model, the penalty on CE can be substantial. This is substantiated by the fact that the CE for the nominal case studied in Figure \ref{fig:ghxall}, with 50\% uncertainty, is already approximately 2.6 and 4.0 times the CE for the LQR applied to the reference system without and with matched disturbances, respectively. A comprehensive use-case will be warranted in which the feasibility assessment of such penalties must be evaluated.

%%%%%%%%%%%%%%%%%%%%%%%%%%%%%%%%%%%%%%%%%%%%%%%%
\section{CONCLUSIONS}
\label{sec:conclusion}
%%%%%%%%%%%%%%%%%%%%%%%%%%%%%%%%%%%%%%%%%%%%%%%%
IES are poised to play a significant role in shaping the future of the American energy landscape. They can enhance the economic competitiveness of critical assets (e.g., SMRs and $\mu$Rs) for reinforcing American energy security, while also supporting the massive deployment of data centers to ensure U.S.~leadership in AI. IES combine electricity generation, heat, and cooling to improve the efficiency of operating each asset separately. However, the heightened complexity of coupled operations, the inherent variability of several asset types (e.g., intermittent energy sources), and the harsh environments that lead to equipment aging introduce uncertainties that make it a challenge to design efficient controllers that coordinate all the involved components.

This study lays the groundwork for using AC to develop a self-regulating controller that adapts in real-time to discrepancies between the expected and actual behavior of the system, which we applied to the TEDS GHX at INL. We first derived a set of assumptions and applied Lyapunov stability analysis to ensure the convergence of tracking errors between the desired reference un-perturbed model and the uncertain system. WE also extended the analysis by demonstrating that in the case of bounded ``matched'' disturbances, the tracking error is bounded. We then evaluated the existing uncertainties in the GHX model and verified that these assumptions hold. Lastly, we deployed AC on the GHX with various levels of uncertainties, both with and without matched disturbances in the form of a sinusoid.

With an uncertainty of 50\%, we found that AC can help decrease the tracking error by 60\%--75\% for $\dot{m}_{ghx,bypass}$ and 32\%--35\% for $Q_{ghx}$ (which was already well tracked by the LR ($x$)). It was also shown to be beneficial to explicitly deal with the matched disturbances in the term $D_{ghx}$ to reduce the MAE and ITAE on the order of 35\% more for $Q_{ghx}$. The improvement aforementioned were achieved despite minimal computing and infrastructure overhead, thus demonstrating its viability and efficacy in the context of real-time control.  However, the CE would increase by up to 4x with matched disturbances and 2.6x without. The next critical task is to deploy these methods on a physical hardware system \cite{jayaram2025model,deghfel2024new}; however, further study of the trade-offs between error tracking and CE will be warranted.

Lastly, AC approaches offer a principled method---grounded in strong mathematical guarantees---for addressing uncertainties. Nevertheless, these approaches are limited by their assumptions. This study qualitatively analyzed the development of a set of mathematical assumptions general enough to apply to the GHX yet rigorous enough to guarantee the boundedness of the states and the convergence of the tracking error. In \cite{seurin2025control} the hours of overhead could be outweighed by considering that the chance constraints approach is less dependent on the amount of uncertainties, while AC require a ``small enough'' uncertainties.  Control represents a vast area of research, and multiple techniques present various advantages for dealing with uncertainties in linear (e.g., LQG \cite{Vajpayee2021l1adaptive}) or nonlinear systems (e.g., sliding mode control \cite{abdulraheem2025aload} and adaptive back-stepping \cite{das2024design,kojic2002adaptive}), GP model form error correction \cite{seurin2025control}, or adaptive observers for partially observable states \cite{Vajpayee2021l1adaptive,annaswamy2023adaptive}. Enhancing the linear formulation derived in this work for more general use cases is a focus of ongoing research.

% ------------------------------------------------------------------------------------
\section*{Credit Authorship Contribution Statement}

 \textbf{Paul Seurin:} Conceptualization, Methodology,  Formal analysis \& investigation, Software, Writing– original draft preparation, Data curation, Visualization, Funding acquisition. \textbf{Anuradha Annaswamy} Methodology, Formal analysis \& investigation, Supervision. \textbf{Linyu Lin:} Investigation, Writing– review \& editing, Data curation. All authors have read and agreed to the published version of the  manuscript.
 
\section*{DATA AVAILABILITY}
The data and codes will, by the time of this publication, be made available upon reasonable demand and in an open-source package hosted on the \href{https://github.com/IdahoLabResearch}{IdahoLabResearch github repository}.

\section*{ACKNOWLEDGMENTS}
This work was supported through the Idaho National Laboratory (INL) Laboratory Directed Research \& Development (LDRD) program under U.S.~Department of Energy (DOE) Idaho Operations Office contract no.~DE-AC07-05ID14517. The authors would also like to thank Dr. Ryan Stewart, Dr. Anand Krishnan, and John Shaver for proof-reading the manuscript.

\appendix
%%%%%%%%%%%%%%%%%%%%%%%%%%%%%%%%%%%%%%%%%
\section{Barbalat's Lemma}
\label{appendix:convergenceofthetrackingerror}
%%%%%%%%%%%%%%%%%%%%%%%%%%%%%%%%%%%%%%%%%
Barbalat's lemma is widely utilized in AC theory; therefore, for the sake of completeness, we recall one variation here.

\textbf{Lemma:} Let $f(t)$ be differentiable; if $\lim_{t\rightarrow \infty} f(t)$ is finite, $f(t)$ is finite, and $\dot{f}(t)$ is uniformly continuous, then $lim_{t\rightarrow \infty} \dot{f}(t) = 0$. A Lyapunov stability certificate can then be obtained for a function $V(x,t)$ such that it satisfies $V(x,t)$ is lower bounded, $\dot{V}(x,t)$ is negative semi-definite $\dot{V}(x,t)$ is uniformly continuous, then $\dot{V(x,t)}\rightarrow 0$ as $t\rightarrow \infty$.

\bibliographystyle{elsarticle-num-names} 
\bibliography{refs}

\end{document}